\documentstyle[12pt,aasms4]{article}

\slugcomment{submitted to the Ap.J.} 

\lefthead{Panagia et al.}
\righthead{Young stars around SN~1987A}

\newcommand{\hst}{\emph{HST}}
\newcommand{\wfpc}{\emph{WFPC2}}
\newcommand{\ha}{H${\alpha}~$}
\newcommand{\ie}{\emph{i.e.}\ }

\newcommand{\eg}{\emph{e.g.}\ }
\newcommand{\cf}{\emph{c.f.}\ }
\newcommand{\mg}{{\it mag}}
\newcommand{\pp}{$^{\prime\prime}~$}

\begin{document}

\title{Young Stellar Populations Around SN~1987A$^\dag$}
\renewcommand{\thefootnote}{\fnsymbol{footnote}}\footnotetext[2]{Based
on observations with the NASA/ESA Hubble Space Telescope, obtained at
the Space Telescope Science Institute, which is operated by
AURA,~Inc., under NASA contract
NAS~5-26555.}\renewcommand{\thefootnote}{\arabic{footnote}} 

\author{Nino Panagia\altaffilmark{1} }
\affil{Space Telescope Science Institute, 3700 San Martin Drive,
Baltimore, MD~21218} 

\author{Martino Romaniello}
\affil{ESO, Karl-Schwarzschild-Stra{\ss}e 2, D-85748 Garching bei
M\"unchen, Germany} 

\author{Salvatore Scuderi}
\affil{Osservatorio Astrofisico di Catania, Viale A. Doria 6, I-95125,
Catania, Italy} 

\and

\author{Robert P. Kirshner}
\affil{Harvard-Smithsonian Center for Astrophysics, 60 Garden Street,
Cambridge, MA~02138} 

\altaffiltext{1}{On assignment from the Astrophysics Division, Space
Science Department of ESA} 

\begin{abstract}
We present the first results of a study of the stellar population
in a region of 30~pc radius around SN~1987A, based on an analysis of
multi-band \emph{HST-WFPC2} images. 

The effective temperature, radius and, possibly, reddening of each
star were determined by fitting the measured broad band magnitudes to
the ones calculated with model atmospheres. In particular, we have
determined effective temperatures and bolometric luminosities for
21,995 stars, and for a sub-sample of 2,510 stars we also determined
individual reddening corrections. In addition, we have identified all
stars with \ha equivalent widths in excess of 8~\AA\,  amounting
to a total of 492 stars. 

An inspection to the HR diagram reveals the presence of several
generations of young stars, with ages between 1 and 150~$Myrs$,
superposed on a much older field population ($0.6-6~Gyrs$). A
substantial fraction of young stars have ages around 12$~Myrs$
which is the stellar generation coeval to SN~1987A progenitor. The
youngest stars in the field appear to be strong-line \emph{T~Tauri}
stars, identified on the basis of their conspicuous ($W_{eq}>8$\AA)
\ha excesses. This constitute the first positive detection of low mass
(about 1-2 $M_\odot$) Pre-Main-Sequence (PMS)  stars outside the Milky
Way. Their positions in the HR diagram appear to require that star 
formation in the LMC occurs with accretion rates about 10 times higher
than in the Milky Way, \ie $\sim10^{-4}M_\odot~yr^{-1}$.

SN~1987A appears to belong to a loose, young cluster 12$\pm$2 Myrs
old, in which the slope of the present mass function is almost identical
to Salpeter's, \ie $\Gamma=dlogN/dlogM\simeq -1.25$ for masses above
$3~M_\odot$, but becomes much flatter for lower masses, \ie
$\Gamma\simeq -0.5$. 

On a large scale, we find that the spatial distribution of massive
stars and low-mass PMS stars are conclusively different, indicating
that different star formation processes operate for high and low mass
stars. This results casts doubts on the validity of an Initial Mass
Function (IMF) concept on a small scale (say, less than 10$pc$).
Moreover, it appears that a determination of the low-mass end IMF in
the LMC requires an explicit identification of PMS stars.  A
preliminary analysis, done for the whole field as a single entity,
shows that the IMF slope for the young population present over the
entire region is steeper than $\Gamma\simeq-1.7$. 

\end{abstract}



\section{Introduction}
SN~1987A has been one of the first targets ever observed with \hst. It
was first imaged with the FOC on August 23-24 1990, immediately
revealing the high potential of \hst\ observations, despite the
unfortunate problem of spherical aberration, with an impressive display
of the supernova itself, flanked by its companion stars, Star~2 and~3,
and the famous inner circumstellar ring (\cite{jak91}, \cite{pan91}). 
The full \hst\ capabilities were reached only after the first
refurbishment mission in December 1993. And surely enough, SN~1987A
caught everybody's eye with its beautiful triple ring circumstellar
nebula which was imaged with the new \wfpc\ (\cite{bur95},
\cite{pan96}). Since then, SN~1987A has regularly been imaged in a
number of broad and narrow band filters at least once a year as part of
the SINS ({\bf S}upernova {\bf IN}tensive {\bf S}tudy, PI: Kirshner)
project to monitor the evolution of the supernova as well as the changes
in the emission of the rings.  The combination of these images, always
centered on the supernova but taken with different roll angles, resulted
in  an excellent coverage of an area of about 130$^{\prime\prime}$
radius, \ie about 30$~pc$, around SN~1987A. 

SN~1987A is located about 20~arcminutes SW of the edge of the 30~Doradus
nebula. This area includes regions of very active star formation, in
which different groups of early type stars are interspersed with HII
regions and SNR shells. The association closest to SN~1987A is LH~90,
which is located about 5 arcminutes to the NE of the supernova
(\cite{luc70}) and whose age is much younger than that of SN~1987A
progenitor, \ie about 4~$Myrs$ as compared with the 10-11~$Myrs$ as
estimated for Sk~-69~202 (\eg \cite{vand98}). The supernova itself is
surrounded by a number of bright blue stars that seem to cluster around
it. It is clear that the study of SN~1987A neighborhood offers a unique
opportunity to place the supernova explosion in the proper context of
stellar evolution and the evolution of stellar populations. 

Here, we present the first results of a systematic study of the stellar
populations contained in a field of 130\pp radius (\ie about 30~$pc$)
around SN~1987A, focusing mostly on the properties of the younger
populations. We show that several distinct generations of young stars
can be identified in addition to the one that gave birth to the SN
progenitor, with ages spanning from 1~$Myr$ or less to at least
100-150~$Myrs$.  Thanks to \ha narrow band photometry, we have
identified almost 500 strong-line T Tauri stars: this constitutes the
first positive (spectroscopic) detection of moderate/low mass (about 1-2
$M_\odot$) Pre-Main-Sequence (PMS) stars outside the Milky Way \footnote
{Existing studies of other young regions (\eg \cite{gil94},
\cite{hun95}) in the LMC have already found PMS stars but only in a
statistical sense, \ie identifying them only on the basis of their
location in the HR diagram.}. Their detection provides an excellent
tracing of the spatial distribution of low mass PMS stars which is
compared with the one of coeval, more massive stars. On this basis, we
show that within the stellar generation of SN~1987A the more massive
stars ($M>6~M_\odot$) are mostly grouped around the SN progenitor
whereas lower mass stars ($1M_\odot<M<2~M_\odot$) are more evenly
distributed, suggesting that different formation mechanisms are
operating for stars of different masses. Finally, we consider and
discuss the problem of determining the Initial Mass Function (IMF),
concluding that an explicit identification of pre-MS stars based on
spectroscopic criteria is absolutely necessary to separate young,
low-mass stars from older population, field stars properly. 

\section{Observations and Data Reduction} 

Since 1994, Supernova~1987A was imaged with the \wfpc\ every year as a
part of the SINS project. The log of the observations we have used is
reported in Table~1. We did not use the observations of March 3, 1995,
since only 4 bands were taken and they did not provide any significant
addition to the portion of sky covered by the other ones. Among all SINS
images taken until 1997, only for one epoch (July~1997) was the field
observed in the \ha\ filter. Since, as shown in Sections 4 and 5, \ha\
images are essential in studying the young population, we complemented
the SINS data with the \ha\ images taken for \hst\ program \#5203 (PI:
John Trauger) on February 3, 1994, which overlap almost exactly with the
SINS images obtained in February 1996 with broad band filters. 


\vskip.1in
\centerline{EDITOR: PLACE TABLE 1 HERE}
\vskip .15in

The observations were processed through the standard PODPS (Post
Observation Data Processing System) pipeline for bias removal and flat
fielding. In all the cases the available images for each filter were
combined to remove cosmic rays events. 

The plate scale of the camera is 0.045 and 0.099 arcsec/pixel in the PC
and in the three WF chips, respectively. We performed aperture
photometry following the prescriptions by Gilmozzi (1990) as refined by
Romaniello (1998), \ie measuring the flux in a circular aperture of
2~pixels radius and the sky background value in an annulus of internal
radius 3~pixels and width 2~pixels. Aperture photometry is perfectly
adequate for our study because crowding is never a problem in our
images. Actually, the average separation of stars from each other is
about 1.3\pp~ (\ie $\sim29$ pixels in the PC chip and $\sim13$ pixels in
the WF one) which is much larger than the \wfpc\ PSF width. The flux
calibration is obtained using the internal calibration of the \wfpc\
(\cite{whi95}), which is typically accurate to within $\pm$ 5\%. We use
the spectrum of Vega as photometric zero point. However, since we have
used the \emph{IRAF-synphot} synthetic photometry package to compute the
theoretical Data Numbers to be compared to the observed ones, the choice
of the zero points has no influence on the final results. 

The brightest stars in each CCD chip are saturated (\ie most stars
brighter than 17.5 in the F555W band). We recover their photometry either
by fitting the unsaturated wings of the PSF for moderately saturated
stars, \ie  with no saturation outside the central 2 pixel radius, or by
following the method developed by Gilliland (1994) for heavily saturated
stars. In most cases, the achieved photometric accuracy is better than
0.05 magnitudes. Full description of the methods used can be found in
Romaniello (1998) and Romaniello {\emph et al.} (1999b). As a sanity
check we have compared our photometry in the B and V bands with the one
obtained with ground based observations (28 stars brighter than
$V\simeq20$ within 30\pp from SN~1987A; Walker and Suntzeff 1990) and
found an overall excellent agreement, \ie {\it rms} deviations less than
0.05 \mg~ in both bands, except for the obvious cases of few stars that
appear as point-like objects in ground based images but are resolved by \hst~
observations. 

Figure~\ref{fig:mos} shows the entire field as observed in a mosaic of
B, V, and R broad band plus the [OIII] and \ha\ narrow band images. As
mentioned in the introduction, SN~1987A appears to be part of a group of
early type stars, whose spatial density is significantly higher in the
neighborhood of the supernova than in the remaining area, suggesting the
presence of a physical group and/or a small cluster. Quantitatively, the
number of stars brighter than $B=18$~\mg\ in the inner
$20^{\prime\prime}$ radius region (15 stars) is about 23\% of the total
64 such stars contained within $130^{\prime\prime}$ radius field,
although the former has an area of only 4\% of the total field. 

\placefigure{fig:mos} 

We identify the stars in the F555W exposure and then measure their
magnitudes in all of the other filters. The total number of stars
detected in this way is 21,955. For about 12,340 of them the photometric
accuracy is better than 0.1~\mg\ in the V, R and I filters. The number
of stars with accuracy better than 0.1~\mg\ drops to 6825 in the B band,
and only 786 stars have a UV filter uncertainty smaller than 0.2~\mg. 

\section{Color-Magnitude Diagrams and HR Diagrams} Figure~\ref{fig:CMD}
displays the Color-Magnitude Diagrams (CMD) for four combinations of
bands. In order to select stars with overall good photometry, we have
used the average error in 5~bands ($\overline{\delta}_5$), excluding the
UV: 

\begin{equation}
\overline{\delta}_5=\sqrt\frac{\delta_{\mathrm{F336W}}^2+
\delta_{\mathrm{F439W}}^2+\delta_{\mathrm{F555W}}^2+
\delta_{\mathrm{F675W}}^2+\delta_{\mathrm{F814W}}^2}{5}
\label{eq:mer}
\end{equation}

\placefigure{fig:CMD}

The black dots in each CMD in Figure~\ref{fig:CMD} are the 6,695 stars 
with $\overline{\delta}_5<0.1$~\mg. The error threshold of
$\overline{\delta}_5<0.1$ reflects itself as a magnitude threshold at
\mbox{$m_{\mathrm{F555W}}\simeq~23$}. We estimate that completeness at
this magnitude limit is very close to 100\% because the density of
detected stars down to 23th magnitude is rather low, \ie about 1 star
per 4.6 square-arcsec area. Thus, conservatively adopting an effective
PSF area of 0.28 square-arcsec (\ie a 3 WF-pixel radius circular area)
and defining $f$ as the probability of having one star brighter than
23rd magnitude falling in any one PSF area ($f=0.28/4.6=0.061$), a
simple application of Poisson statistics predicts that only a fraction
of $[1-fe^{-f}/(1-e^{-f})] = 0.031$ of the stars may overlap with other
stars. 

It is apparent that, despite the high quality of the measurements
(internal uncertainties less than 0.1 magnitudes), the various features
of the CMDs, such as the Zero Age Main Sequence (ZAMS) and the sequence
of binaries for the early type stars, and the red giant clump for the
more evolved populations, are rather ``fuzzy" and not sharply defined. 
Although this is due in part to the presence of several stellar
populations projected on each other (see Sections 4-6), most of the
problem arises from the fact that reddening is not quite uniform over
the field, thus causing the points of otherwise identical stars to fall
in appreciably different locations of the CMDs. 

The large number of bands available (6 broad band filters) which cover a
wide baseline (more than a factor of 4 in wavelength, extending from
$\sim$2,300 to $\sim$9,600~\AA) provide us with a sort of {\it wide-band
spectroscopy} which defines the continuum spectral emission distribution
of each star quite well.  Therefore, by comparison with model
atmospheres (\cite{bes98}), one can fit the 6 band observations of each
stars and solve for 3 unknowns simultaneously, namely the effective
temperature, $T_{eff}$, the reddening, $E(B-V)$, and the angular radius,
$R/D$. As well known, the solution in the plane $T_{eff}-E(B-V)$ may not
be unique for stars with effective temperatures lower than about
9,000~$K$. Therefore, we first solve for the full set of parameters,
$T_{eff}$, $E(B-V)$, and $R/D$ only for stars suitably selected on the
basis of reddening-free colors. These turn out to have temperature
either higher than 10,000~$K$ or between 6,750 and 8,500~$K$. For each
of the remaining stars, we adopt the average reddening of its neighbors
and solve for only two parameters, $T_{eff}$, and $R/D$. Finally the
stellar luminosity is computed from the derived $T_{eff}$ and $R/D$
values, adopting a distance to SN~1987A of 51.4~$kpc$ (\cite{pan99},
\cite{pana99}, \cite{romadist99}). A full account of this analysis will
be presented in a forthcoming paper (\cite{roma99}; see also
\cite{rom98}). Out of the total sample of 21,995 stars, the uncertainty
in the effective temperature is lower than 12\% for 9,474 stars, most of
which with $log(L/L_\odot)>0$. Also, we were able to obtain individual
reddening determinations for 2,510 stars (\ie about 11\% of the whole
sample). On average, this corresponds to a reddening value every 13
square arcseconds, which implies that the extinction distribution over
the entire field was mapped down to a scale size of about 3.5\pp (\ie
about 1~$pc$). 

The reddening is clearly not constant over the field, having an average
value of $<E(B-V)>=0.203$ and a {\it rms} deviation
$\sigma(E(B-V))=0.072$. The average value is very close to the reddening
determined by Scuderi {\emph et al.} (1996) in the direction of Star~2
($0.19\pm 0.02$) and by Walborn {\emph et al.} (1993) from an analysis
of photometric ground based observations of early type stars in the same
field. The reddening spatial distribution does not correlate well with
the distribution of nebular emission.  Moreover, the reddening
fluctuations do not show any obvious pattern, suggesting the presence of
a highly clumped diffuse medium (\cf \cite{rom98}, \cite{roma99}). 

\section{Stellar Populations and Ages} The resulting HR diagram
($\log(L/L_\odot)~vs.~\log(T_{eff})$ plot) is shown in
Figure~\ref{fig:hr}. Its inspection confirms the early findings of
Walker and Suntzeff (1990) and Walborn {\emph et al.} (1993) and reveals
that the distribution of stars in the HR diagram is clearly bound toward
high temperatures, identifying a ZAMS that corresponds to a metallicity
$Z\simeq Z_\odot/3$ (\cf Figure 3a). This value agrees well with the
general abundances measured in the LMC as well as the more local
estimates derived from spectroscopy of the circumstellar rings
(\cite{pan96}) and Star~2 (\cite{scu96}). Therefore, in the following
the observed points will be compared with $Z=0.006$ isochrones, either
from Brocato \& Castellani (1993) and Cassisi, Castellani \& Straniero
(1994) for post-MS evolution or from Siess et al (1997) for PMS
evolution. 

\placefigure{fig:hr} 

With the exception of one very bright star\footnote{This star has
$log(T_{eff})=4.7, ~log(L/L_\odot)=5.8$, which correspond to a mass of
$\sim60~M_\odot$ and suggest an age probably younger than a million
years.} (readily recognizable in Figure 1 and denoted with a larger
symbol in Figures 3b and 6c), the positions of the most luminous blue
stars (see Figure 3b) fall on isochrones corresponding to ages around
10-14~$Myrs$, which make them coeval to SN~1987A progenitor and Star~2
(\cf \cite{scu96} and references therein). There are a number of stars
at intermediate luminosities and temperatures that indicate ages up to
100-150~$Myrs$ and initial masses down to 4~$M_\odot$. This is
especially clear from the position of the yellow/red supergiants around
$\log(T_{eff})\sim3.6-3.8$ and $\log(L/L_\odot)\sim 2-3$. 

The lower MS and the Red Giants are mostly old population stars,
consistent with a metallicity either identical to, or slightly lower
than the one of the young components.  No single age can explain the
distribution of the old population, and substantial star formation
between 600~$Myrs$ and 6~$Gyrs$ is required to account for the
observations (\cf Figure 3b). 

In summary, we find evidence for a series of star formation episodes
spanning from several billion years ago until at least 1-2~$Myrs$ and,
possibly, still ongoing (\cf Figure 4 and Section 5). We identify a
generation of stars with ages about 12~$Myrs$ which is coeval to the
supernova progenitor and may include up to 35\% of the young population
in this field (defined as stars with ages $<100~Myrs$). The old field
population requires continued star formation over a broad interval of
$0.6-6~Gyrs$. From all of these results, it is clear that, since even
the young stars belong to a mixture of populations of different ages, a
study of the IMF is very hard and requires a proper separation of the
various stellar generations to avoid systematic biases and errors. 

\section{Pre-Main-Sequence Stars and the Young Population } 

By comparing the magnitudes in the R band (F675W) with the ones measured
with the narrow band H$\alpha$ filter (F656N) we can identify stars with
sufficiently strong H$\alpha$ emission. Considering the throughputs of
the F675W and F656N filters (\cf {\it WFPC2 Instrument Handbook},
\cite{wfpc2_hb}) we estimate that an H$\alpha$ emission line with
equivalent width of 8~\AA\ will produce a color excess
$m(R)-m(H\alpha)\simeq0.3$~\mg. Therefore, we define as stars with
strong H$\alpha$ excess the ones with a color excess $m(R)-m(H\alpha)$
which is both greater than 0.3 and greater than 4 times the photometric
uncertainty, \ie $>4\sqrt{\sigma^2(R)+ \sigma^2(\mathrm{H}\alpha}$). The
requirement on the error assures that a measured excess is highly
significant, and the high equivalent width cutoff excludes any
contamination by normal stars with chromospheric activity, because they
have equivalent widths of less than 3~\AA\ (\eg \cite{fra94}). The
positions in the HR of \ha excess stars are shown in Figure 4 and
compared with the distribution of all other stars. 

\placefigure{fig:bepms} 

We identify the luminous and bright ones (4 stars), which are near the
MS, as Be stars. One of them is Star 3, one of the two companions to
SN~1987A (see \cite{walb93} and references therein): its high
variability qualifies it for a ``typical" Be star (\cf \cite{jas80}).
However, its emission lines are too narrow (\cf \cite{walb93}) to
conform to the characteristics of canonical Be stars. A possible
explanation is that Star 3 is seen almost pole-on.  On the other hand,
we cannot exclude that Star 3 belongs to the same class of HAeBe stars,
\ie massive PMS stars (Herbig 1960), as the six possible candidates
found near the center of the LMC bar by Lamers, Beaulieu and De Wit
(1999). They concluded that the location in the HR diagram of their
candidate HAeBe stars indicate a prevailing accretion rate of about
$10^{-4}~M_\odot~yr^{-1}$, \ie ten times higher than the ``canonical
value" of $10^{-5}~M_\odot~yr^{-1}$ as appropriate for the Milky Way
(\cf \cite{pal93}). Also our 4 Be stars, if they are considered to be
stars evolving toward the Main Sequence, would require a considerably
brighter birthline for the LMC than it is found for Milky Way stars (see
Fig 4), again indicating accretion rates up to
$10^{-4}~M_\odot~yr^{-1}$. If the PMS nature of such stars is confirmed
by detailed spectroscopic studies, these results would imply that star
formation in a low metallicity environment proceed at a substantial
different pace than it does in our Galaxies, at least for massive stars.

\placefigure{fig:hahisto} 

We identify the redder and fainter stars with strong \ha excess (488
stars, 275 of which have very accurate temperature determinations,
$\delta\log T_{eff}<0.05$) as strong-line T~Tauri stars, \ie PMS stars
with circumstellar material, remnant of their proto-stellar
cocoons\footnote{Confirmation of the reality of the measured \ha
excesses and the T Tau identification is provided by both a visual
inspection to the images, which excludes the presence of nebular
filamentary structures crossing each candidate star image, that may
mimic an apparent ``excess", and measurements of the stellar fluxes in
the [OIII] and [NII] narrow band filters (F502N and F658N, respectively)
that for all of the {\it bona fide} T Tau candidate stars give line
emission that are appropriately much lower than the emission measured
for the \ha line. }. Their \ha excesses correspond to \ha equivalent
widths which range between 8 and 360 \AA, with 90\% of the
W$_{eq}(H\alpha)$ values falling within 11 and 85~\AA. These properties
are very similar to the ones of galactic strong-line T Tau stars (\eg
\cite{her88} \cite{fern95}) although we note that the \ha equivalent
width distribution of T Tau stars in our field is more skewed toward
lower values than the galactic ones (\cf Fig. 5). This difference is
easily explained as an age effect, in that classical galactic T Tau
stars are much younger (ages below 5 Myrs) than the ones in our field.
Comparing the positions of T Tauri stars in the HR diagram with PMS
evolution model calculations by Siess {\emph et al.} (1997), we find
that most of them have masses in the range 1-2$M_\odot$ and ages between
1-2~$Myrs$ up to 20~$Myrs$ and possibly beyond (see Figure 4). Note that
the presence of T Tau stars on, or very close to the birthline even
suggests that some star formation is still ongoing in this field.
Overall, T Tau stars represent about 4\% of the total number of stars
present in the same area of the HR diagram (approximately,
$log(T_{eff})=3.65-3.95$ and $log(L/L_\odot)<1$), making them very hard
to identify without positive spectroscopic diagnostics. 

We like to stress that this result constitutes the first positive
detection of moderate/low mass (about 1-2 $M_\odot$) PMS stars outside
the Milky Way. In fact, although previous studies of other young regions
of the LMC (\eg NGC~1850 by Gilmozzi {\emph et al.} 1994, R136 cluster
by Hunter {\emph et al.} 1995, {\it etc.}) have already identified low
mass ``PMS stars", those identifications were entirely circumstantial in
that they were based exclusively on the location of the candidate stars
in the HR diagram. In the following we will show that, because of heavy
contamination by older, field populations of low mass stars, such an
``identification" method may, at best, have just a statistical value and
may lead to an overestimate of the actual number of low mass stars in a
star forming region. 
 
In Figure~6a and 6b we display the spatial distribution of massive
stars, \ie stars with M$>$6~M$_{\odot}$, and strong T~Tauri stars as
defined above, overlaid on a mosaic of the \ha\ images.  We see that
massive stars are strongly concentrated near SN~1987A (14 out of 55 are
within 20\pp of the supernova), and the remaining ones are mostly
located East of SN~1987A (31 out of 41). The PMS stars, on the other
hand, do not show any strong spatial concentration, although one can
notice that the number density of PMS stars on the NE side of SN~1987A
is appreciably lower than on the SW side. No obvious correlation is seen
between either class of stars and the ionized gas features: this may
indicate that the region is rapidly loosing memory of the star formation
process once the most massive stars, say, $M>20M_\odot$, have evolved
and died producing SNII explosions. 

\placefigure{fig:space} 

From Figure 6c, which displays the distributions of the two classes of
stars in one and the same plot, it is apparent that they are distributed
in substantially different manners. To better quantify this statement,
we have divided the field into 9 regions: a circle of 20\pp radius
around SN~1987A and 8 radial sectors, as shown in Figure 6d. 

\placetable{tab:sn87a_nd}

Table~\ref{tab:sn87a_nd} summarizes the areas of the 9 regions (note
that for each sector the area reported in Table 2 corresponds to the
area for which {\it both} broad band {\it and} \ha\ observations are
available) their star number counts and the corresponding star densities
per unit area. An inspection to this table reveals that there are
enormous variations in the proportion of massive stars relative to low
mass stars: in particular, the numbers of massive and low mass stars in
the central area are almost the same, while in other areas the low mass
stars are 7 to 28 times more numerous than massive stars. These
differences are highly significant because they greatly exceed the
expected Poissonian fluctuations, also given in
Table~\ref{tab:sn87a_nd}. An experimental check of this fact is provided
by similar statistics on the number of stars belonging to the so-called
Red Giant Clump, \ie stars of a much older population which, therefore,
are expected to be uniformly distributed over the field. And indeed, as
seen in Table~\ref{tab:sn87a_nd}, the observed number densities for Red
Giant Clump stars show fluctuations which are perfectly compatible with
Poisson statistics. 

Another point worth noting is that the average density of T~Tau stars in
sectors 2+3 (\ie East of SN~1987A) is (12.9$\pm$1.4), \ie 3.5$\sigma$
lower than the average 18.6$\pm$0.8, whereas in sectors 4+5 (South of
SN~1987A) is 3.9$\sigma$ higher (27.4$\pm$2.1). It would be tempting to
interpret such a systematic difference in terms of a star formation
``wave" or ``front" that sweeps region. On the other hand, we do not
find any significant difference in ages among PMS stars in the various
sectors, suggesting that, rather than dealing with some sort of
propagating star formation, the key factor here is the overall
efficiency of the star formation process that varies from place to
place. Also, we note that there is no enhancement in the density of
T~Tau stars around the SN~1987A cluster, nor near the brightest star in
the observed field (see Figure 6c) lending support to the idea that the
process of formation of low mass stars may be distinct from the one
leading to the formation of massive stars. 

\section{SN~1987A Stellar Group: a ``Typical" Cluster?} 

In Section 4 we have argued that massive stars are strongly concentrated
 around SN~1987A.  This is clearly seen in Figure 7a (see also Fig. 1)
that shows an area of about 50\pp$\times$50\pp around SN~1987A: 14 stars
brighter than $10^3~L/L_\odot$ are found in the vicinities of the
supernova. Actually, SN~1987A and its companions, Star~2 and Star~3, are
at the Eastern edge of the stellar group, whose center almost coincides
with a compact ``core" of 8 blue stars. We note that the size of this
stellar group associated with SN~1987A (about 9~pc) is similar to the
average size of stellar clusters in the LMC (7.7$\pm1.5~pc$;
\cite{hodg88}) suggesting that we are dealing with a ``canonical" LMC
cluster. And, indeed, this group has been  classified as a ``loose
cluster'' by Kontizas {\emph et al.} (1988; their No. 80). 

\placefigure{fig:cluster} 

Walborn {\emph et al.} (1993) discussed the properties of the cluster
stars brighter than $V\simeq20$.  Analyzing their ground based {\it UBV}
photometry, and adopting an average reddening of $E(B-V)=0.18$, they
concluded that the cluster had an age of about 12$\pm$4 million years. 

As clearly seen in Figure 7b, with our data we can reduce the
uncertainty and assign an age of 12$\pm$2 Myrs to the bulk of the stars
present in this field. This age can fairly well account for the
positions in the HR diagram of both the most massive stars (with one
exception; see below) {\it and} the PMS stars found in this field. There
is one bright star ($log~T_{eff}\simeq 4.59$, $log~L/L_\odot \simeq
4.51$) that appears to lie on the ZAMS and, therefore, to have a much
younger age, say, 5 Myrs or less.  While statistical errors cannot
account for this discrepancy, we notice that this star displays a
significant \ha excess ($W_{eq}\simeq 12$\AA) and, therefore, it is one
of the 4 Be stars we mentioned in section 4.  We argue that the strong
Be characteristics are enough to alter the photospheric properties of
this star so as to make it {\it appear} hotter and brighter than its
average parameters would allow.  In particular, if this is an almost
pole-on Be star, we should {\it expect} its color temperature to be
significantly higher than average. A ``natural" consequence, which
suggest a simple test to run, is that for a pole-on Be star the width of
the \ha emission is expected to be relatively narrow. 

Given the number of stars in the mass range 6-20~$M_\odot$ (15 stars
including the supernova progenitor), and adopting an IMF with slope
$\Gamma$ in the range -1 to -2, we estimate that the cluster originally
included additional 2$\pm$1 stars in the mass range 20-80~$M_\odot$,
which exploded as SNII well before the SN~1987A event. However, there is
no clear sign of any shell-like structure centered on the cluster. This
implies that either the upper mass cutoff for {\it this} cluster was
much lower than normal, \ie no star more massive than $\sim20~M_\odot$
formed in this cluster, or the more massive stars gave origin to
explosions involving considerably lower kinetic energies than the
``canonical" value of $10^{51}~erg$. 

The supernova  is not at the center of the cluster but rather it is
offset about 15\pp to the East relative to the massive star barycenter,
(\ie the center of mass of 14 stars more massive than 6$M_\odot$ plus Sk
-69 202 itself). We speculate that this lack of ``prominence" for the
supernova progenitor may indicate that the few cluster stars, which were
more massive than the SN~1987A progenitor and exploded long ago, were
distributed over the cluster body, so as to fill the ``gaps" between
either the SN-Star2-Star3 group, or the SE cluster extension (4 massive
stars), and the cluster center (8 massive stars).

\section{Initial Mass Function: A Discussion} 

As pointed out in Section 5, the number ratio of PMS low mass (1-2
$M_\odot$) stars to massive stars in the central cluster is much smaller
than the average in the whole field, \ie 1.4 as compared to the average
value 8.9, indicating that there is an objective deficiency of low mass
stars in this young cluster. Such a deficiency would translate into an
IMF slope that over the approximate mass range $1-14~M_\odot$ is
considerably flatter than in  the general field. In particular, if one
adopts a slope $\Gamma = dlogN/dlogM\simeq -1.5$  for the entire field
(see next Section) the slope in the compact cluster would be
approximately $\Gamma=-0.5$. On the other hand, a different conclusion
would be reached if one limits the analysis to relatively high mass
stars.  For example, if one assumes all stars brighter than
$100L/L_\odot$ to belong to one and the same stellar generation, the IMF
between 3 and 14 $M_\odot$ would turn out to have a slope $\Gamma\simeq
-1.2$, \ie very similar to Salpeter's IMF. From these results one might
conclude that a single power-law cannot represent the observations, and
that the IMF agrees with Salpeter's for masses higher than $3M_\odot$
and flattens dramatically for lower masses. 

On the other hand, it is rather incautious to talk about an IMF when
dealing with small number statistics, and/or with very limited regions
in space. In addition, looking at the cluster HR diagram (\cf Fig. 7b)
one can notice that many PMS stars appear to be appreciably older than
10 Myrs, and actually appear to cluster around a 20 Myrs isochrone. This
fact, while confirming that the most recent stellar  generation is poor
of low mass stars, also tells us that the formation of lower mass stars
in the cluster region is occurring on different timescales than that of
more massive stars. 

On a larger scale, the almost anti-correlation of the spatial
distributions of high mass and low mass stars of a coeval generation
(\cf Section 5, and Fig. 6) indicates that star formation processes for
different ranges of stellar masses are rather different and/or require
different initial conditions. An important corollary of this result is
that the very concept of an ``initial mass function" (IMF) may not have
validity in detail, but rather be the result of a chaotic process, so
that it may make sense to talk about an {\it average IMF} over a
suitably large area in which all different star formation processes are
concurrently operating. Actually, if we just take the ratio of the total
numbers of massive to low-mass stars belonging to the young population
as reported in Table~\ref{tab:sn87a_nd}, and interpret them in terms of
a power-law IMF adopting mass intervals of 6-15~$M_\odot$ and
1-2~$M\odot$ for massive stars and for PMS stars, respectively, we would
derive a slope of the initial mass function of
$\Gamma=dlogN/dlogM\simeq-1.3$ with a purely statistical uncertainty of
$\pm0.1$. Such a value is remarkably close to the classical Salpeter's
(1955) slope $\Gamma=-1.35$ for the Solar neighborhood and is valid over
essentially the same mass interval as the original Salpeter's analysis,
\ie $\sim$2-10~M$_\odot$. However, this is a conclusion drawn without
taking into account of possible incompleteness effects. On the one hand,
for both massive and  low mass stars above $\sim1M_\odot$,
incompleteness due to missed detection and/or to crowding/blending  is a
negligible effect because all of these stars are well above our
detection limit and their surface density in not so high (1 star per
$\sim170$ WF pixel area, or, equivalently, an average separation of
stars of $\sim13$ WF pixels).  On the other hand, for PMS stars one has
bear in mind that we can reliably identify only strong-line T~Tau stars
and that the {\it total} number of PMS stars of comparable masses may be
considerably larger. For example, Alcal\'a {\emph et al.} (1996) have
shown that in the Orion region the number of weak-line T Tau stars is at
least comparable to, and possibly larger than that of strong-line T
Tauri stars. Actually, if one would identify PMS stars on the basis of
their UV excess, \ie as stars with much bluer $U-B$ colors than expected
for normal stars with their observed $B-V$ and $V-I$ colors, one would
count in our field as many as $\sim850$ PMS stars (\cite{rom98}), \ie a
factor of 1.7 more than the positively identified T~Tau stars. Moreover,
T~Tau stars are known to exhibit strong H$\alpha$ variability on short
timescales (\eg Smith {\emph et al.} 1999, and references therein). As a
consequence, at any one time one may be able to detect only a fraction
of the entire population of strong-line T~Tau stars. Indeed, preliminary
comparisons of overlapping regions in fields imaged at two different
epochs (about 15\% of the entire field around SN~1987A) have shown that,
while the number of strong-line T~Tau stars identified at any one time
is essentially constant, no more than half of the candidate T~Tau stars
display a significantly strong H$\alpha$ excess at both epochs
(\cite{romp99}). This suggests that the total number of PMS stars may
easily be twice as high as the number of strong-line T~Tau stars we have
identified, and possibly even higher. In turn, this result implies a
significantly steeper IMF slope, say, $\Gamma\simeq-1.7$ or even more
negative.  A detailed discussion of these effects is presented in
Romaniello (1998). Eventually, combining observations of the SN~1987A
field taken at substantially more epochs than available at present, we
will be able to properly identify {\it all} T Tau stars in the field so
as to obtain a reliable statistics of the young population in the
neighborhood of SN~ 1987A and to reach firm conclusions about star
formation processes and history in this region of the LMC
(\cite{romp99}). 

It is important to realize how crucial it is to {\it individually}
identify and characterize each and everyone of the PMS stars if one
wants to evaluate an IMF reliably.  For example, in our field we can
identify about 500 PMS stars using an H$\alpha$ excess criterion, and
possibly almost 1000 on the basis of a UV excess. Still, such numbers
represent a small fraction of the {\it total} number of stars with
temperatures and luminosities comparable to those of the candidate PMS
stars. Actually, limiting ourselves to stars which in the HR diagram
fall between the 5 and 40~$Myrs$ PMS isochrones  and which are brighter
than $L_\odot$ (\ie for stars whose statistics is pretty complete), the
ratio of the total number of stars to the number of {\it bona fide} PMS
stars is $\sim20$ if we include all possible candidates, and more than
40 if we include only strong-line T~Tau stars.  This means that in the
absence of an explicit characterization of PMS stars, using just the
location of stars in the HR diagram  as the criterion to recognize PMS
stars will unavoidably introduce some heavy contamination. One can argue
that, for a fixed surface density of old population stars, such an effect
is much reduced when studying regions containing large concentrations of
young stars. For instance, the possible contamination by older
populations could  be, say, 10\% or less, if the surface density of
young stars in a given region were higher than 200-400 times the density
of young stars around SN~1987A or, equivalently, 10 times the density of
old population stars in the SN~1987A vicinities. However, since the
density of old population stars with luminosities in the range
$1<L/L_\odot<10$ is about 0.2 stars per square arcsecond, exceeding that
density by a factor of 10,  at least, would imply an average density of
young population, low mass stars ($1<L/L_\odot<10$) higher than 2 stars
per square arcsecond, and at least ten times higher for stars in the
next 1-$dex$ bin in luminosity. It is easy to realize that with such
high stellar densities another problem is bound to arise, at least for
low mass stars, namely confusion/blending  due to crowding. These
effects will both ``drown" faint stars into a ``sea" of even fainter
stars and/or would artificially create brighter stars by confusing
nearby stars into one more luminous, apparently point-like source
(Panagia 2000). 

From this discussion, it follows that it is essential to have {\it
spectroscopic} criteria that allow one to discern PMS stars from field
stars unambiguously and completely.  H$\alpha$ excess and/or UV excess
are possible ways of accomplishing this goal, but even these methods
need confirmation, calibration and sharpening, in that we still have to
compare our multi-band photometry (or ``wide-band spectroscopy") with
{\it real} spectra  before a 100\% reliable of identification of T Tau
stars can be claimed.  We are going to fill this gap by taking
medium-high resolution spectra of a number of our best T Tau candidates
with the ESO-VLT1 in January 2000 (\cite{fava00}). With that it will
eventually be possible to start talking about low mass star formation
and low end of the IMF in the LMC on firm scientific grounds.

\section{Conclusions} 

We have analyzed the \wfpc~ images obtained in the UV, U, B, V, R, I
broad bands, plus the [OIII] and H$\alpha$ narrow band filters, of the
130\pp radius region centered on SN~1987A, focusing on the properties of
the stellar populations present in the field.  The main results can be
summarized as follows: 

\begin{itemize} 

\item Over the entire field, we identify 21,995 stars, of which 6,695
have an overall photometric accuracy in the optical bands better than
0.1\mg. 

\item From a comparison of the available 6 broad band photometry with
model atmospheres, we have determined effective temperatures and
bolometric luminosities for all of the 21,995 stars.  For a subset of
2,510 we also determined  individual reddening corrections.  Overall, we
obtained an excellent characterization of a large fraction of the sample
stars. In particular, the effective temperature was determined with an
uncertainty lower than 0.05~$dex$ in $log(T_{eff})$ for 9,474 stars,
most of which are brighter than L/L$_\odot$=1. 

\item From an analysis of the resulting HR diagram we have found
evidence for a series of star formation episodes spanning from several
billion years ago until at least 1-2~$Myrs$ and, possibly, still
ongoing. 

\item We have identified a generation of stars with ages about
10-12~$Myrs$ which is coeval to the supernova progenitor and may include
up to 35\% of the young population in the entire field (defined as stars
with ages $<100~Myrs$). 

\item The properties of the field old population indicates strong and
continued star formation over a broad interval of $0.6-6~Gyrs$. 

\item We have identified 488 strong-line \emph{T~Tauri} stars on the
basis of their conspicuous ($W_{eq}>8$\AA) H$\alpha$ excesses. Their
ages peak at about 10-20 $Myrs$ and range from about 1 $Myrs$ up to
possibly 40 $Myrs$.  Their positions in the HR diagram appear to require
that star formation in the LMC occurs with accretion rates about 10
times higher than in the Milky Way, \ie $\sim10^{-4}M_\odot~yr^{-1}$.
This constitute the first positive detection of low mass (about 1-2
$M_\odot$) PMS stars outside the Milky Way.

\item SN~1987A appears to belong to a loose, young cluster 12$\pm$2 Myrs
old, in which the slope of the present mass function is almost identical
to Salpeter's, \ie $\Gamma\simeq -1.25$ for masses above $3~M_\odot$,
but becomes much flatter for lower masses, \ie $\Gamma\simeq -0.5$. 

\item A comparison of the spatial distributions of massive stars and
low-mass PMS stars shows that  they are conclusively different, which
indicate  that different star formation processes operate for high and
low mass stars. 

\item We have addressed the problem of determining an IMF in our field.
Because of the substantial differences of the spatial distributions of
stars of different masses in our field, we conclude that the very
concept of an IMF is not valid on a small scale (say, $< 10~pc$
diameter, \ie a scale comparable to the size of SN~1987A cluster) but it
may become meaningful on a larger scale. A preliminary analysis shows
that, averaging over the whole region (about 60$pc$ diameter), the IMF
slope is steeper than Salpeter's IMF, \ie $\Gamma\simeq-1.7$.

\item Since even the young stellar population is a mixture of
generations of stars with different ages, we have concluded that a study
of the IMF requires a proper separation of {\it all} of the various
stellar populations coexisting with each other.  We have shown that a
proper characterization of the low mass end of the IMF requires explicit
identification of individual PMS stars. 

\end{itemize}

\acknowledgments
SS acknowledge the kind hospitality of STScI, were most of this work
was done, as well partial support from the STScI Visiting Scientist
Program. The comments of an anonymous referee were valuable to improve 
the presentation. This work was supported in part by HST-STScI grants
GO-6437, GO-7434 GO-7821, GO-8243 to the SINS project (PI: RPK) and by
STScI-DDRF grants \# 82131, 82160, and 82186 to NP. 

\eject 

\appendix

\begin{table*}[!ht]   
\begin{center}
\caption{Log of the observations centered on 
Supernova~1987A.}
\begin{tabular}{*{5}{c}}
& & & & \\
Filter Name &  \multicolumn{3}{c}{Exposure Time (seconds)}  & Comments\\
\cline{2-4} &September 1994\tablenotemark{a}
&February 1996\tablenotemark{b}
&July 1997\tablenotemark{c}   & \\ \tableline
{\bf F255W} &      2x900      &    1100+1400   &   2x1300    & UV~Filter  \\
{\bf F336W} &      2x600      &      2x600     &   2x800     & ~U~Filter  \\
{\bf F439W} &      2x400      &     350+600    &   2x400     & ~B~Filter  \\
{\bf F555W} &      2x300      &      2x300     &   2x300     & ~V~Filter  \\
{\bf F675W} &      2x300      &      2x300     &   2x300     & ~R~Filter  \\
{\bf F814W} &      2x300      &      2x300     &   2x400     & ~I~Filter
\\[0.2cm]\tableline & & & & \\[-0.4cm]
{\bf F502N} &     4x1200      &    1100+1500   &2x1300+4x1400&[OIII] $\lambda$
5007~\AA \\
{\bf F656N} &      ---        & 1100+1300\tablenotemark{d} &  4x1400 & \ha \\
{\bf F658N} &     4x1200      &    1100+1500   &     ---     & [NII] $\lambda$
6584~\AA\tablenotemark{e} \\
\end{tabular}
\end{center}

\tablenotetext{a}{September 24, 1994, proposal number 5753.}
\tablenotetext{b}{February 6, 1996, proposal number 6020.}
\tablenotetext{c}{July 10, 1997, except for F502N taken on
July 12, 1997, proposal number 6437.}
\tablenotetext{d}{Not from SINS (PI John Trauger), taken on February 
3, 1994.}
\tablenotetext{e}{With $\sim$25\% \ha contamination.} 
\label{tab:log}
\end{table*}

\begin{table*}[!ht]
\caption{{\bf Spatial distribution of stars around SN1987A} -
Observed numbers and derived densities of the various types of stars
in the regions illustrated in Figure~4d. The quoted errors are the
expected fluctuations according to Poisson statistics.} 
\begin{minipage}{\linewidth}
\begin{center}
\begin{tabular}{*{8}{c}}
 & & & & & & & \\
\hline
Region & Area\protect{\footnote{In units of the area of the central cluster
(20\pp radius, \ie 1207 square arcseconds).}} &
\multicolumn{2}{c}{Massive stars\protect{\footnote{$\log(\mathrm{L/L}_\odot)
>3$, $M>6$~M$_\odot$.}}} &
\multicolumn{2}{c}{T~Tauri stars\protect{\footnote{$EW(\mathrm{H}\alpha)
\gtrsim8$~\AA, 1~M$_\odot<M<2$~M$_\odot$.}}}&
\multicolumn{2}{c}{Red Giants\protect{\footnote{$3.65<\log(\mathrm{T}_{eff})
<3.9\
\mathrm{and}\ 1.5<\log{\mathrm{L/L}_\odot}<2.1$, LMC field population.}}} \\
\cline{3-8}
 &  & Number & Density & Number & Density & Number & Density \\ \hline
 & & & & & & & \\[-0.2cm] 
C & 1.00 & 14 & $14\pm3.7$  & 20 & $20.0\pm4.5$ & 16 & $16.0\pm4.0$ \\

1 & 4.70 & 10 & $2.1\pm0.7$ & 81 & $17.2\pm1.9$ & 62 & $13.2\pm1.7$ \\

2 & 3.80 &  9 & $2.4\pm0.8$ & 50 & $13.2\pm1.9$ & 48 & $12.6\pm1.8$ \\

3 & 3.16 &  6 & $1.9\pm0.8$ & 40 & $12.6\pm2.0$ & 34 & $10.7\pm1.8$ \\

4 & 3.02 &  6 & $2.0\pm0.8$ & 83 & $27.5\pm3.0$ & 40 & $13.2\pm2.1$ \\

5 & 3.14 &  3 & $1.0\pm0.6$ & 84 & $26.8\pm2.9$ & 57 & $18.2\pm2.4$ \\

6 & 2.31 &  0 &     0       & 38 & $16.4\pm2.7$ & 37 & $16.0\pm2.6$ \\

7 & 2.41 &  3 & $1.2\pm0.7$ & 43 & $17.8\pm2.7$ & 38 & $15.8\pm2.6$ \\

8 & 2.69 &  4 & $1.5\pm0.7$ & 49 & $18.2\pm2.6$ & 25 & $ 9.3\pm1.9$ \\

\hline
 & & & & & & & \\ 
Total & 26.23 & 55 & $2.1\pm0.3$ & 488 & $18.6\pm0.8$ &
       357 & $13.6\pm0.7$ \\
\hline\hline
\end{tabular}
\end{center}
\end{minipage}
\label{tab:sn87a_nd}
\end{table*}

\clearpage

%
%

\clearpage

\figcaption{The field centered on SN1987A (about 130\pp~ radius) as
observed in the combination of the B, V, and I broad bands plus the
[OIII] and H$\alpha$ narrow band images. The blue, bright star about
94\pp NE of SN~1987A is the brightest ($log(L/L_\odot)\simeq5.8$) and
most massive ($\sim 60M_\odot$) star in the field (see Sections 4 and
5). 
\label{fig:mos}} 

\figcaption{Color-Magnitude Diagrams for four combination of filters.
The grey dots are stars with average errors (as defined in Eq. (1))
$\overline{\delta}_5>0.1$, whereas the black dots are the 6,695 stars
with $\overline{\delta}_5<0.1$. 
 \label{fig:CMD}} 

\figcaption{The HR diagram for the stars in the \wfpc\ field. The grey
dots are stars with average errors (as defined in Eq. (1))
$\overline{\delta}_5>0.1$, whereas the black dots are the 6695 stars
with $\overline{\delta}_5<0.1$. {\bf [a]}~{\it (left panel)}: Comparison
with the theoretical ZAMS for different Z values: $Z=Z_\odot$ long
dashed line, $Z=0.3Z_\odot$ solid line, and $Z=Z_\odot/20$ dotted line.
~{\bf [b]}~{\it (right panel)}: Comparison with a number of post-MS
isochrones (\cite{bro93}, \cite{cas94}). The brightest star in the
field, possibly a $\sim 60M_\odot$ star younger than 1 $Myrs$, is marked
with a circle at the top of the diagram. 
\label{fig:hr}} 

\figcaption{The HR diagram displaying the positions of the stars with
strong \ha excess (dots) overlayed on the general stellar population
(grey squares) found in the \wfpc\ field. For reference, we show
the theoretical ZAMS (\cite{bro93}) with marked position for stars of
various masses.  Also shown are 2.5-20~$Myrs$ PMS isochrones
(\cite{sie97}) and the birthlines for accretion rates of $10^{-4}$ 
(short-dashed line) and $10^{-5}~M_\odot~yr^{-1}$ (long-dashed line;
\cite{pal93}). 
\label{fig:bepms}} 

\figcaption{The distribution of number of stars with strong \ha
emission per unit of $0.2~dex$ bin of the \ha equivalent width,
$W_{eq}(H\alpha)$. Candidate Be stars have been excluded from this
histogram.  The shaded histogram is the corresponding distribution for
galactic strong-line T Tau stars (\cite{fern95}). 
\label{fig:hahisto}} 

\figcaption{Comparison of the spatial distributions of massive stars
($M>6M_\odot$, star symbols) and PMS stars ($M<2M_\odot$, squares)
belonging to the same younger population.  North is up and East is to
the left.  {\bf [a]}~{\it (upper left)}: Massive stars overlayed on a
mosaic of \wfpc ~H$\alpha$ images. {\bf [b]}~{\it (upper right)}: PMS
stars overlayed on a mosaic of \wfpc ~H$\alpha$ images. {\bf [c]}~{\it
(lower left)}: Massive stars (blue star symbols) and PMS stars (red
square symbols) overlayed on the general field including 21,995 stars
(little dots). The brightest star ($log(L/L_\odot)\simeq5.8$) is denoted
with a larger size, pale blue star symbol. {\bf [d]}~{\it (lower
right)}: Number of massive stars (blue), PMS stars (red) and Red Giant
Clump stars (black) in a central circle centered on SN~1987A and in 8
sectors (\cf~ Table and Section 5).
\label{fig:space}} 

\figcaption{The properties of the central cluster: {\it Left panel -}
The combined BVR plus \ha image of an area 50\pp$\times$50\pp around
SN~1987A. {\it Right panel - } The corresponding HR diagram in which
stars with strong \ha excess are denoted with red symbols. Isochrones
for post-MS evolution at 10 and 14 Myrs, and PMS evolution at 5, 10 and
20 Myrs are also shown. 
\label{fig:cluster}} 

\end{document}